\documentclass[prl,twocolumn,superscriptaddress,noshowpacs,notitlepage,longbibliography]{revtex4-1}%
\usepackage{graphicx,bm,times}
\usepackage{amsmath}
\usepackage{amsfonts}
\usepackage{amssymb}
\usepackage{color}
\usepackage{hyperref}
\hypersetup{
	colorlinks = true,
	allcolors = {blue}
}

\begin{document}
	\title{Revealing the single electron pocket of FeSe in a single orthorhombic domain}
	
	\author{Luke C. Rhodes}
	\thanks{These two authors contributed equally.}
	\affiliation{School of Physics and Astronomy, University of St. Andrews, St. Andrews KY16 9SS, United Kingdom}	
	
	\author{Matthew D. Watson}
	\thanks{These two authors contributed equally.}
	\affiliation{Diamond Light Source, Harwell Campus, Didcot, OX11 0DE, United Kingdom}
	
	\author{Amir A. Haghighirad}
	\affiliation{Institute for Quantum Materials and Technologies, Karlsruhe Institite of Technology, 76021 Karlsruhe, Germany}
	
	\author{Daniil~V.~Evtushinsky}
	\affiliation{Laboratory for Quantum Magnetism, Institute of Physics, \'{E}cole Polytechnique F\'{e}d\'{e}rale de Lausanne, CH-1015 Lausanne, Switzerland}
	
	\author{Timur K. Kim}
	\affiliation{Diamond Light Source, Harwell Campus, Didcot, OX11 0DE, United Kingdom}
	
	\begin{abstract}
		We measure the electronic structure of FeSe from within individual orthorhombic domains. Enabled by an angle-resolved photoemission spectroscopy beamline with a highly focused beamspot (nano-ARPES), we identify clear stripe-like orthorhombic domains in FeSe with a length scale of approximately 1-5~$\mu$m. Our photoemission measurements of the Fermi surface and band structure within individual domains reveal a single electron pocket at the Brillouin zone corner. This result provides clear evidence for a one-electron pocket electronic structure of FeSe, observed without the application of uniaxial strain, and calls for further theoretical insight into this unusual Fermi surface topology. Our results also showcase the potential of nano-ARPES for the study of correlated materials with local domain structures.
	\end{abstract}
	\date{\today}
	\maketitle
	
	
	Electronic nematicity, a correlated electronic state which breaks rotational symmetry, is a common phenomenon in many strongly correlated materials, including cuprates \cite{Daou2010}, heavy Fermion systems \cite{Okazaki2011} and iron-based superconductors \cite{Kasahara2012}. In the iron-based superconductors, nematicity manifests as a tetragonal to orthorhombic structural phase transition which, despite a very small change in lattice constants, coincides with large anisotropic responses in measurements that are sensitive to the material's electronic structure \cite{Fisher2011}. It has been postulated that this nematic state may be directly related to the superconducting properties of these materials \cite{Fernandes2014}, however, the proximity of the nematic phase to stripe-like antiferromagnetic ordering in most compounds means that determining the key properties of a pure nematic state has proved challenging.
	
	FeSe is an ideal system for investigating the origin of the nematic state as it exhibits a well-defined nematic transition at  $T_s = 90$~K  without the presence of long range magnetic order \cite{Pustovit2016,Coldea2018,BoehmerKreiselReview2017}. The electronic structure of FeSe has been extensively investigated by angle-resolved photoemission spectroscopy (ARPES) and significant changes in the band structure can be observed to onset at the nematic transition \cite{Nakayama2014,Watson2015,Watson2016,Fedorov2016,Fanfarillo2016,Watson2017}. However, the typical size of the beam spot used for high-resolution synchrotron ARPES measurements is on the order of 50~$\mu$m or larger \cite{Hoesch2017}, while individual orthorhombic domains can be on the order of only 1-10~$\mu{}$m \cite{Tanatar2015}. Therefore conventional ARPES measurements will usually detect a superposition from both domains with approximately equal intensity.
	
	Several groups have attempted to overcome this limitation by aligning the orthorhombic domains along a preferential axis by applying uniaxial strain \cite{Watson2017b,Yi2019,Huh2020,Pfau2019}, commonly known as ``detwinning". These measurements revealed a single elliptical hole-like pocket at the Brillouin zone centre, in line with theoretical predictions \cite{Mukherjee2015}. However, they also revealed the presence of only a single electron-like pocket around the M and A points at the Brillouin zone corner, despite the expectation of two electron pockets in most theoretical models \cite{Mukherjee2015,Watson2016,Fanfarillo2016,Kang2018}. The theoretical understanding of this discrepancy remains controversial and unresolved \cite{Rhodes2018,Kushnirenko2018,Christensen2020,Yi2019,Benfatto2018,Kreisel2017}.

	\begin{figure*}
		\centering
		\includegraphics[width = \linewidth]{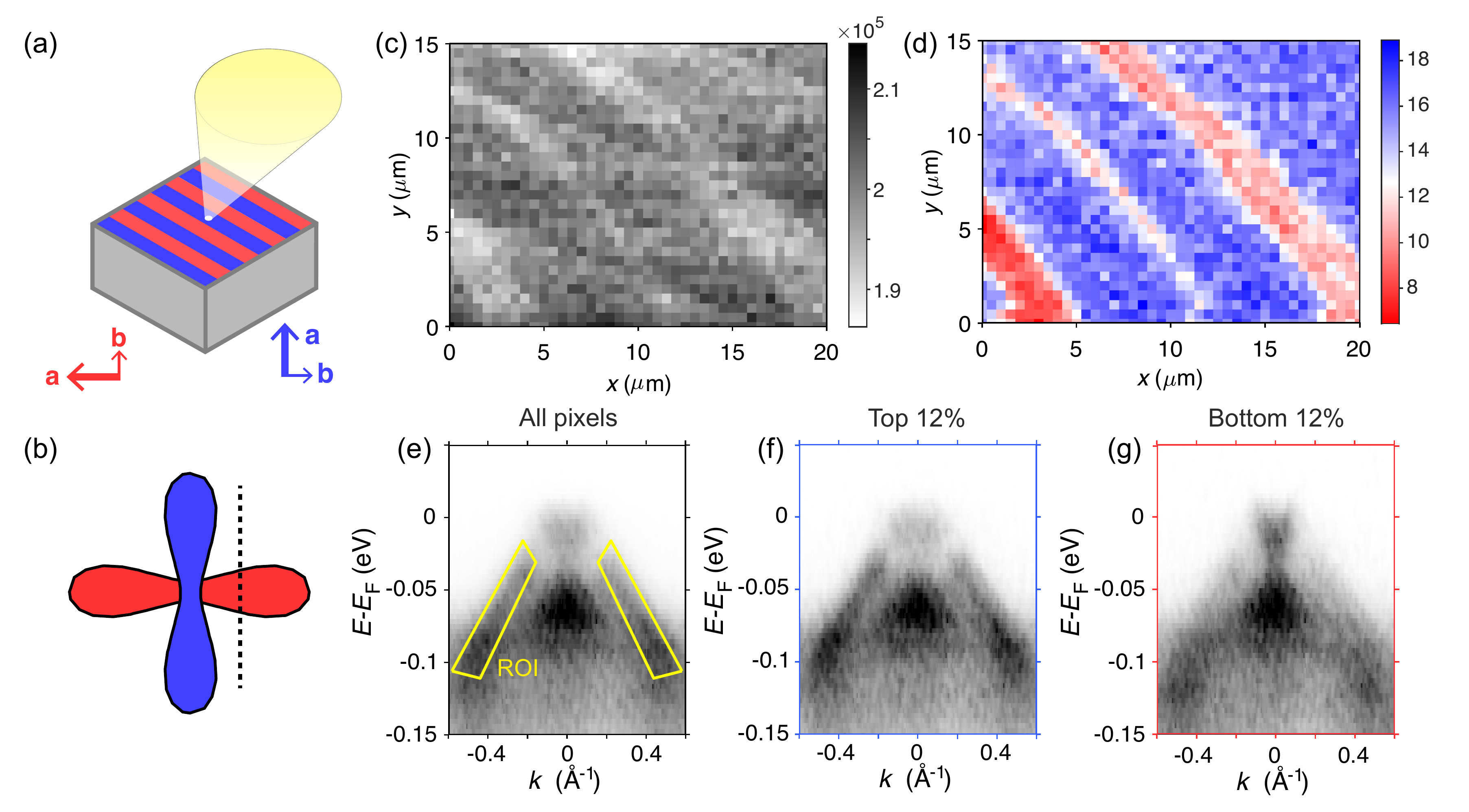}
		\caption{Domain structure of FeSe. a) Schematic of the experiment. Red and blue stripes represent domains with 90$^\circ$-rotated orthorhombic axes. Nano-ARPES measurements focus the photon beam to around 1$\mu$m$^2$ which allows the detection of photoelectrons from a single domain only. b) Sketch of the cut in momentum space, just off the high symmetry M-point, used in this figure to uncover the domain structure. In a one-electron-pocket scenario, the band dispersions along this cut will include an electron band crossing the Fermi level in one domain only. c) Total photoelectron intensity real-space map of a region of cleaved surface of FeSe, taken at $h\nu$ = 58 eV with 0.5$\mu$m steps. d) The same map as (c) but with the photoelectron intensity integrated only within the region of interest (ROI) shown in (e). e) The energy dispersions obtained by the summation of all pixels. The regions of interest used in (d) are highlighted in yellow. f,g) Energy dispersions obtained by summing the top 12\% of pixels from (d) (i.e. from blue areas) and the bottom 12\% of pixels (from red areas).}
		\label{fig:fig1}
	\end{figure*}
	
	There is, however, a more favourable approach to study the electronic structure of FeSe. Rather than applying uniaxial strain to the sample, which may alter the materials physical properties or deform its surface, one can focus a photon beam spot below the typical length scale of the domains. This constitutes a significant technical challenge, requiring the combination of good spatial, angular, and energy resolutions, with a photon energy and geometry set to reach the more intriguing bands near the Brillouin zone corner, whilst also maintaining a reasonable count rate and keeping the sample position and temperature under control. However, recent progress in nano-ARPES makes such experiments now achievable \cite{Iwasawa2019,Watson2019}, and here we report the first such measurements of the nematic state of FeSe.
	
	In this Letter, we use nano-ARPES to measure the electronic structure of FeSe domain-by-domain. We begin by detecting regular stripe-like orthorhombic domains with a periods of 1-5~$\mu$m. By measuring the electronic structure from within a single domain, we reveal that the states around the M-point are consistent with the one-electron pocket description observed by ARPES measurements on detwinned crystals. This result provides direct and apparent evidence for a one-electron pocket at the M-point in FeSe, and calls for deeper theoretical insight into the origin of the electronic nematic state. Our measurements showcase the potential for measuring single-domain electronic structure in correlated materials using the technique of nano-ARPES.
	
\begin{figure}
		\centering
		\includegraphics[width=\linewidth]{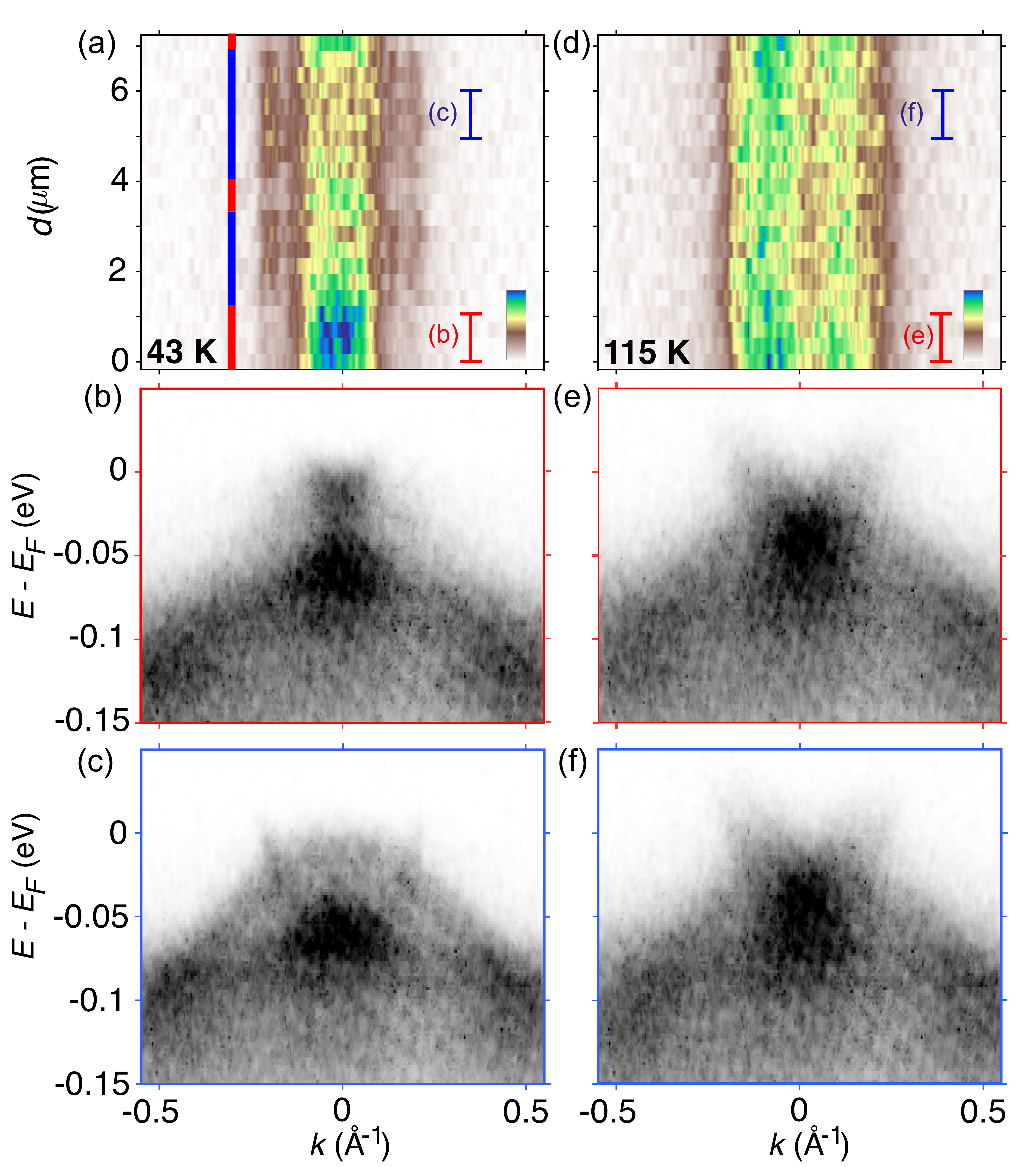}
		\caption{Line scan across the domains. a) Photoemission intensity integrated within 10 meV of $E_F$ as a function of spatial distance along a spatial cut perpendicular to the domains in Fig. \ref{fig:fig1}(d). In $k$-space, the cut corresponds to the A-Z direction. The red and blue striped line is a guide to indicate the different orthorhombic domains. b,c) Energy dispersions averaged over the area defined by the red and blue profiles in (a). d-f) Equivalent measurements at $T$=115~K, where the lack of spatial dependence highlights the loss of any domain structure above $T_s$.}
		\label{fig:fig2}
	\end{figure}

	Single crystal were grown using the chemical vapor transport method \cite{Boehmer2014PRB,Watson2015}. Nano-ARPES measurements were performed at the I05 beamline at the Diamond Light Source, UK \cite{Hoesch2017}. Fresnel zone plate technology is used to focus the beam. While the beamline is capable of spatial resolutions better than $\sim$500~nm \cite{Rosner2019,Iwasawa2019}, our measurements are performed with a beamspot (full-width half maximum) closer to 1~$\mu{}$m, as a compromise to achieve a higher count rate. The beamline is equipped with a Scienta DA30 analyser. We use a photon energy of 58~eV in linear vertical polarisation, chosen to access the A-point where the electron bands are slightly better separated than at M \cite{Watson2016}. The total energy resolution was approximately 30 meV and the angular resolution was 0.3$^\circ$. Thus, the energy resolution is not comparable to the highest-resolution ARPES measurements on FeSe \cite{Rhodes2018,Watson2016,Liu2018} as some compromise is necessary to achieve a reasonable count rate; nevertheless it is sufficient to resolve the bands with clarity. The sample temperature was set to 43~K throughout the duration of the experiment, except for the measurements of the tetragonal state (Fig. \ref{fig:fig2}(d-f)) at 115~K.

	In order to establish the presence of a domain structure, we must begin by analysing spatial mapping data. As shown schematically in Fig.~\ref{fig:fig1}(a) it is necessary that our beam spot must be smaller than the typical domain spacing. We choose a geometry described in Fig.~\ref{fig:fig1}(b), where, according to the one-electron pocket scenario, only the ``red" domain should show an electron pocket at the Fermi level.  In Fig. \ref{fig:fig1}(c) we show a spatial map of the surface of FeSe, where the colormap of each pixel corresponds to the total number of photoelectrons in the detector. This hints at some underlying domain structure, but the variation in total intensity is only on the order of 10$\%$; in other words, the two domain orientations give a similar but not identical number of total photoelectrons in this measurement geometry. However, the true domain structure can be much more clearly revealed by integrating the counts detected within a particular region of interest (ROI) in momentum and energy space. We present this map in Fig. \ref{fig:fig1}(d) where we have used two polygon-shaped ROIs defined by Fig. \ref{fig:fig1}(e). Using a divergent red-blue colorscale for emphasis, we can identify multiple stripe-like domains, with a typical length scale of 1-5~$\mu$m. In our geometry, the spatial $x$ and $y$ axes correspond approximately to the Fe-Fe bond directions, i.e.the domain boundaries are at 45$^\circ$ to the Fe-Fe directions, in keeping with previous results on other iron-based superconductors \cite{Fisher2011,Tanatar2015}.
	
	To clarify that we are sensitive to the different band dispersions in the two domains, we perform a filtering procedure on this 4-dimensional mapping data to take the spatial pixels which contain either the top 12\% intensity within the ROI or the bottom 12\%, and sum over these pixels to produce the energy dispersions in Fig.~\ref{fig:fig1}(f) and \ref{fig:fig1}(g). The top 12\% of pixels from Fig. \ref{fig:fig1}(d), corresponding to the ``blue" domains, shows intense hole like bands extending towards, but not reaching, the Fermi level. In contrast, the bottom 12\%, corresponding to the ``red" domains, reveals a small bright electron pocket, in accordance with the expectation of only one domain giving a Fermi-crossing band in this geometry \cite{systematicissue}.

		Now that we have established the domain structure of our sample, we can study the band dispersions through the high symmetry A-point. To do this, we have taken 45$^\circ$ line scans, i.e. equal steps in $x$ and $y$, bisecting the orthorhombic domains shown in Fig. \ref{fig:fig1}(d). The intensity measured around the Fermi level in presented in Fig. \ref{fig:fig2}(a). In the ``red" domains, bright intensity can be seen with a small $k_F$ of $\sim$0.05~\AA, which is consistent with the $k_F$ value of the minor length of the electron pocket determined by high resolution ARPES \cite{Watson2016}. In contrast, the ``blue" domain shows less intensity within the $\sim$0.1~\AA~region (there is still a hole-like band present near $E_F$ so there is some central intensity), but instead presents a large $k_F$ value of $\sim$0.2 \AA, again consistent with published values for the $k_F$ value of the longer axis of the electron pocket at A. By averaging over a few points of the line scan that are well within individual domains, as indicated in Fig.~\ref{fig:fig2}(a), we achieve the effective single-domain high-symmetry energy dispersions presented in Fig. \ref{fig:fig2}(b) and \ref{fig:fig2}(c). In Fig. \ref{fig:fig2}(b) a small electron pocket can be observed with a minimum near $E_F$, and a large dispersion broadly centered around -50~meV, whereas in Fig. \ref{fig:fig2}(c) a large electron-like band dispersion can be observed, which exhibits a minimum around -50~meV, and an additional hole-like dispersion which extends up to just below $E_F$. In both dispersions, a broad band centered just below -50~meV can be detected. Overall the spectra closely resemble ARPES measurements on detwinned crystals \cite{Watson2017b}.
		
		At 115~K, above the nematic transition, tetragonal symmetry is restored. Thus, in Fig.~\ref{fig:fig2}(d) the $k_F$ value as a function of spatial distance now appear constant, and in Fig.~\ref{fig:fig2}(e,f) we observe identical band dispersions, indicating an absence of any significant spatial variation. These spectra are in agreement with dispersions observed at the the A-point in the tetragonal phase \cite{Watson2016,Rhodes2017}, and indicate the presence of two electron pockets above $T_s$.
		
		\begin{figure*}
			\centering
			\includegraphics[width=0.8\linewidth]{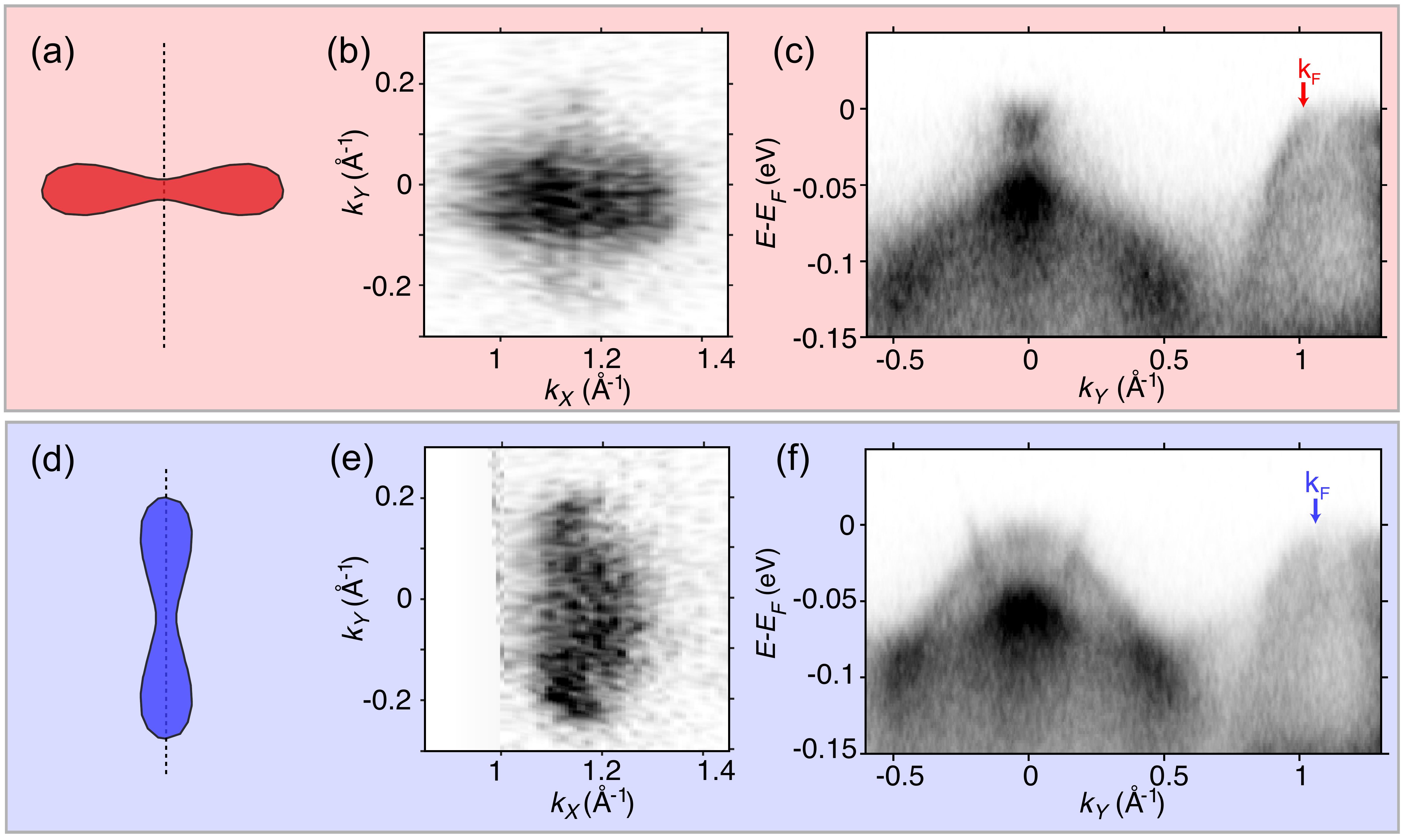}
			\caption{Fermi surface around the A-point and A-Z band dispersion from two orthorhombic domains. a) Schematic of the single electron pocket Fermi surface at the A-point. b) Measured Fermi surface map highlighting a single electron pocket elongated along the experimental $k_X$ axis. c) A-Z band dispersion along the experimental $k_Y$ axis, as shown by the dashed line in (a). d-f) Equivalent measurements from an adjacent domain with 90$^\circ$-rotated orthorhombic axes. Here we use the notation $k_X$ and $k_Y$ to refer to the momentum with respect to the fixed experimental geometry \cite{Yi2019,Watson2017b}, where the analyser angle corresponds to $k_Y$.}
			\label{fig:fig3}
		\end{figure*}
		
		Finally, we present Fermi surface maps and high statistics dispersions obtained from static positions, precisely chosen to be well within each of the domains. The Fermi surface maps around the A-point (Fig. \ref{fig:fig3}(b,e)), are obtained using the DA30 analyser's deflector-mode mapping while the sample position remains fixed. In either case, a single pocket can be observed at the Fermi level, elongated along either the $k_X$ or $k_Y$ direction, and drawn schematically in Fig. \ref{fig:fig3}(a,d).
		We further present high-statistics measurements in Fig. \ref{fig:fig3}(c,f), obtained at the same spatial positions, which extend the electronic structure measurements from the A-point towards the Z-point of the second Brillouin zone. It can be seen that there is also some observable difference here: a larger $k_F$ value of the hole pocket is observed in Fig. \ref{fig:fig3}(c) compared to Fig. \ref{fig:fig3}(f), confirming that the elongation of the hole pocket is perpendicular to that of the electron pocket \cite{Suzuki2015,Watson2017b}. Overall, the dispersions closely resemble previous photoemission measurements on detwinned crystals of FeSe \cite{Watson2017b,Yi2019,Huh2020,Pfau2019}, though it should be noted that the geometry here is not fully equivalent to some reports. We do not comment on the orbital characters of the bands, which has been strongly disputed \cite{Christensen2020,Watson2016,Pfau2019}, but only note that the connectivity of bands from the Z to the A points (or $\Gamma$ and M) is not obvious in such measurements due to the weak and broad features in between.
		
		
		The length scales we observe correspond to those obtained by polarised light imaging \cite{Tanatar2015}, i.e. extensive stripe-like domains with widths on the order of 1-5~$\mu{}$m. Interestingly, the blue domain appears to be preferential within the 20$\times{}$15~$\mu$m area presented in Fig.~\ref{fig:fig1}(d), and presumably the red domain would dominate elsewhere on the sample. Such variation on a more macroscopic length scale was observed by the laser micro-ARPES measurements of Schwier \textit{et. al.} \cite{Schwier2019}, and was understood to correspond to local strain variation favouring particular domain orientations, rather than individual domains. We propose, therefore, that the resolution of the characteristic stripe-like domains in spatial mapping data is the preferred method to justify any claim of obtaining ARPES spectra that is truly from a ``single domain" \cite{Liu2018,Hashimoto2018}.
		
		Within the context of the debate on the electronic structure of FeSe, our nano-ARPES results are consistent with those obtained by detwinning the sample, and we were unable to find any quantitative difference in band positions.
This provides direct evidence that uniaxial strain has a negligible effect on the materials intrinsic electronic structure, at least up to the point where the sample is almost fully detwinned. Thus, the four detwinned ARPES measurements \cite{Watson2017b,Yi2019,Huh2020,Pfau2019} along with this nano-ARPES experiment should solidify the one-electron pocket Fermi surface of FeSe as the true nematic ground state of this material.
		
		Such a dramatic shift between a two-electron pocket A-point at 100~K to a one-electron pocket at low temperatures provides a natural explanation for many of the anisotropic properties detected in FeSe at low temperatures, including the resistivity anisotropy \cite{Tanatar2015}, strongly anisotropic spin susceptibility revealed by inelastic neutron scattering \cite{Chen2019} and the unusual twofold-symmetric momentum dependence of the superconducting gap \cite{Sprau2017,Rhodes2018}. It has also been shown that such an electronic structure is fully consistent with quasiparticle interference measurements \cite{Rhodes2019} and seems consistent with the measured Sommerfeld coefficient of FeSe \cite{Watson2015,Hardy2018}. Despite the mounting experimental evidence, however, it remains challenging to identify a reasonable theoretical model of the electronic structure of FeSe which can describe the evolution of the Fermi surface from a two-electron pocket tetragonal state to a one-electron pocket nematic state. Discovering such a model would be a key step forward for understanding the origin and importance of nematicity within FeSe and potentially all iron-based superconductors. Recent theoretical calculations have suggested that a hybridisation between the  $d_{xz}$ and $d_{xy}$ states may be an important factor \cite{Long2019_arXiv,Christensen2020}.
		
		In a wider context, nano-ARPES is a novel and fast-improving technique that has already had significant impact in the field of 2D materials, especially in heterostructures and devices \cite{Nguyen2019,Ulstrup2019,Cattelan2018Review}.  However, thus far it has seen only relatively few applications in traditional correlated electron systems \cite{Iwasawa2019,Watson2019}, at least partially because the lower energy scales involved require high energy resolution, which is challenging to combine with micron-scale spatial resolution. Nevertheless, in this work we were able to measure nematic domains in FeSe, where the length scale can be as short as a 1-2~$\mu{}$m and the energy scales on the order of 20~meV. Since the formation of magnetic or structural domains is common in correlated systems, our results highlight the tremendous opportunities for the use of nano-ARPES to reveal new structure in correlated electronic phases.
		
		\begin{acknowledgments}
			We thank Cephise Cacho for technical support during this experiment. We acknowledge Diamond Light Source for time on beamline I05-ARPES under proposal SI23890. LCR acknowledges funding from the Royal Commission for the Exhibition of 1851.
		\end{acknowledgments}
		

\begin{thebibliography}{45}
\bibitem [{sys()}]{systematicissue}%
	\BibitemOpen
	\href@noop {} {}\bibinfo {note} {Our filtering procedure leads to a
		systematic bias towards finding intense hole-like bands inside the ROI in
		Fig.~\ref{fig:fig1}(f); however there is no systematic bias towards the
		absence of an electron-like pocket in Fig.~\ref{fig:fig1}(f) but its strong
		presence in Fig.~\ref{fig:fig1}(g), which further confirms that we indeed
		resolve the domain structure.}\BibitemShut {Stop}%
\end{thebibliography}


%

\end{document}